\documentclass[prl,aps,amssymb,twocolumn,showpacs,superscriptaddress]{revtex4}
\usepackage{graphicx}
\usepackage{amsmath}

\begin{document}

\title{Shaping a time-dependent excitation to control the electron distribution function:\\
noise minimization in a tunnel junction}

\author{Julien Gabelli}
\affiliation{Laboratoire de Physique des Solides, Univ. Paris-Sud, CNRS, UMR 8502, F-91405 Orsay Cedex, France}
\author{Bertrand Reulet}
\affiliation{Laboratoire de Physique des Solides, Univ. Paris-Sud, CNRS, UMR 8502, F-91405 Orsay Cedex, France}
\affiliation{Universit\'e de Sherbrooke, Sherbrooke, Qu\'ebec J1K 2R1, Canada}

\date{\today}
\begin{abstract}
 We report measurements of shot noise in a tunnel junction under bi-harmonic illumination, $V_{ac}(t)=V_{ac1} \cos(2\pi \nu t) + V_{ac2} \cos(4\pi \nu t+ \varphi)$. The experiment is performed in the quantum regime, $h\nu \gg k_BT$ at low temperature $T = 70 \, \mathrm{mK}$ and high frequency $\nu = 10 \, \mathrm{GHz}$. From the measurement of noise at low frequency, we show that we can infer and control the non-equilibrium electronic distribution function by adjusting the amplitudes and phase of the excitation, thus modeling its shape. In particular, we observe that the noise depends not only on the amplitude of the two sine waves but also on their relative phase, due to coherent emission/absorption of photons at different frequencies. By shaping the excitation we can minimize the noise of the junction, which no longer reaches its minimum at zero dc bias. We show that adding an excitation at frequency $2\nu$ with the proper amplitude and phase can \emph{reduce} the noise of the junction excited at frequency $\nu$ only.
\end{abstract}

 \maketitle

\vspace{-0.5cm}
In recent years the dynamical control of mesoscopic conductors has gained increasing interest, mainly motivated by the realization of a phase-coherent electronics for quantum computation. One major challenge is the experimental achievement of a single electron excitation above the Fermi sea. The way to drive the ground state of a metallic conductor to reach the single electron excitation should be optimized to minimize the creation of electron-hole excitations; this can be probed by noise measurements \cite{Levitov1,Levitov2,Klich,Degiovanni,Ambrumenil,Vanevic}. For a conductor with energy-independent transmission, the variance of the current fluctuations due to the discrete nature of electrons, the so-called shot noise, reaches a minimum when the excitation $V_{L}(t)$ is a $T=1/\nu$ periodic sequence of Lorentzian peaks of quantized area $\int_0^T eV_L(t) \, dt=Nh$ each, with $N$ integer. In a tunnel junction this leads to a noise spectral density $S_2=Ne^2\nu$, i.e. the same as the shot noise of a purely dc current $I=Ne\nu$. Thus, this ac excitation creates a non-equilibrium electron distribution function with the remarkable property that it yields to a charge transfer of $N$ electrons per cycle in average with a variance $\Delta N^2=N$.

It is experimentally difficult to generate Lorentzian pulses with precise shape and high repetion rate, a condition necessary to observe the predictions \cite{Levitov1,Levitov2}. The simplest ac excitation consist of a pure sine wave $V_{ac}\cos (2\pi \nu t)$. Unfortunately, the presence of the ac voltage always increases the noise \cite{Lesovik,Schoelkopf,Reydellet,GR1}, i.e. $S_2(V_{dc},V_{ac})>S_2(V_{dc},V_{ac}=0)$, where $S_2$ is the noise spectral density measured at low frequency. A much richer waveform, which we have used in the present work, is the bi-harmonic excitation:
\begin{equation}
V_{ac}(t)=V_{ac1} \cos(2\pi \nu t) + V_{ac2} \cos(4\pi \nu t+ \varphi)
\label{eqbiharm}
\end{equation}
By controlling the three parameters $V_{ac1}, V_{ac2}, \varphi$, one can modify the shape of the ac excitation, which modifies the out-of-equilibrium electron distribution function and thus the noise. As we show below, adding the excitation at frequency $2\nu$ may lower the noise, i.e. $S_2(V_{dc},V_{ac1},V_{ac2})<S_2(V_{dc},V_{ac1},V_{ac2}=0)$, thus partially erasing the extra noise created by the excitation at frequency $\nu$. This occurs because the absorption/emission of two photons of frequency $\nu$ may interfere destructively with that of one photon of frequency $2\nu$.

 The Letter is organized as follows. (i) We describe the experimental setup. (ii) We calculate the non-equilibrium stationary distribution function generated by any time-dependent, periodic excitation. (iii) From the noise measurements we deduce the experimental electron distribution function in the presence of the bi-harmonic excitation. (iv) We show that a bi-harmonic excitation with two spectral components $\nu$ and $2\nu$ can reduce the mono-harmonic photon-assisted noise at frequency $\nu$.
\begin{figure}[h]
\begin{center}
\includegraphics[width=0.75\linewidth]{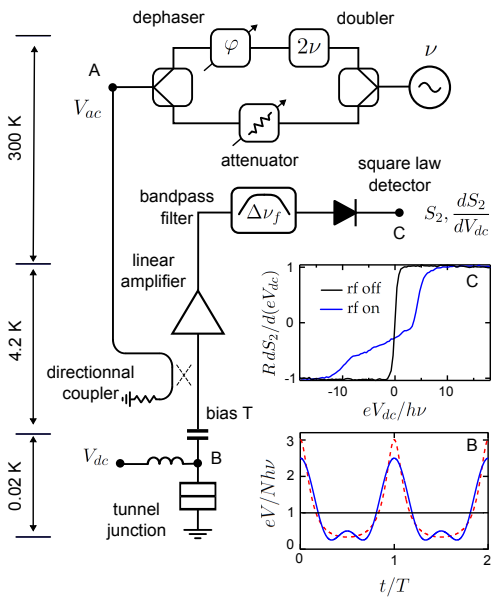}
\end{center}
\vspace{-0.5cm}
\caption{Experimental setup for the measurement of the photon-assisted noise in a tunnel junction under bi-harmonic excitation. Inset B: $T$-Periodic sequence of bi-harmonic excitation (red line) with $eV_{dc}=eV_{ac1}=2eV_{ac2}=h\nu$ and $\varphi=0$, approximating Lorentzian pulses of width $\tau=\mathrm{ln}\,2/(2 \pi \, T)$ and height $N=1$ (red dashed line). Inset C: normalized differential noise spectral density with and without microwave excitation \textit{vs.} normalized dc bias. The power of the generator, the variable attenuator and the phase shifter are tuned to obtain $eV_{ac1}=2eV_{ac2}=5.4h\nu$ and $\varphi=0$.\label{fig1}}
\end{figure}

\emph{Experimental setup.}  We have measured the shot noise of an Al/Al oxide/Al tunnel junction similar to that used for noise thermometry \cite{Lafe} cooled to $70 \,\mathrm{mK}$. We apply a 0.1 T perpendicular magnetic field to turn the Al normal. We measure the noise at low frequency while the junction is excited by the bi-harmonic ac voltage (\ref{eqbiharm}), as depicted in Fig.\ref{fig1}. To generate the bi-harmonic signal, a microwave source of frequency $\nu=10 \, \mathrm{GHz}$ is split in two arms. A frequency doubler in the upper arm generates the oscillating voltage at $2\nu=20 \, \mathrm{GHz}$. Its phase $\varphi$ can be tuned by a mechanical phase shifter while its amplitude $V_{ac2}$ is set by the tunable output power of the source. In the lower arm, a variable attenuator allows to modify $V_{ac1}$. The signals from the two arms are recombined at point A and sent to the sample through a directional coupler placed at liquid helium temperature. A bias tee, sketched by an inductor and a capacitor in Fig.\ref{fig1}, allows to add the dc voltage $V_{dc}$ to the ac one coming from the coupler. An example of an achievable waveform is shown on Fig. \ref{fig1}B, together with a Lorentzian. The ac voltages experienced by the sample are measured by fitting the data of the photo-assisted noise with a single frequency, as in \cite{GR1}. The resistance of the sample $1/G=48 \, \Omega$ is close enough to $50 \, \Omega$ to provide a good impedance matching to the coaxial cable and avoid reflection of the ac excitation. Thus, only the fluctuating current due to the tunneling process is amplified by a low noise cryogenic amplifier (noise temperature $T_N \simeq 7 \,\mathrm{K}$). The noise is filtered to keep frequencies in the range $0.5 - 1.8$ GHz before impinging  on a power detector. The dc voltage at point C is proportional to the noise power density $S_2$ integrated over the bandwidth of the filter. The derivative of the  noise $\partial S_2/ \partial eV_{dc}$ is measured with an additional $77 \, \mathrm{Hz}$, small voltage modulation and a usual lock-in detection. Fig.\ref{fig1}C shows measurements of the differential noise $\partial S_2/ \partial eV_{dc}$ with and without ac excitation. From the data without ac excitation and taking into account the finite bandwidth of the detection, we determine the electrons temperature: $T_{el} = 70 \, \mathrm{mK}= 0.14 \, h\nu/k_B$. When the ac excitation is switched on, the differential noise exhibits an intermediate step, a consequence of the electron energy distribution function differing from that of  Fermi-Dirac.

\begin{figure}[b]
\begin{center}
\includegraphics[width=0.8\linewidth]{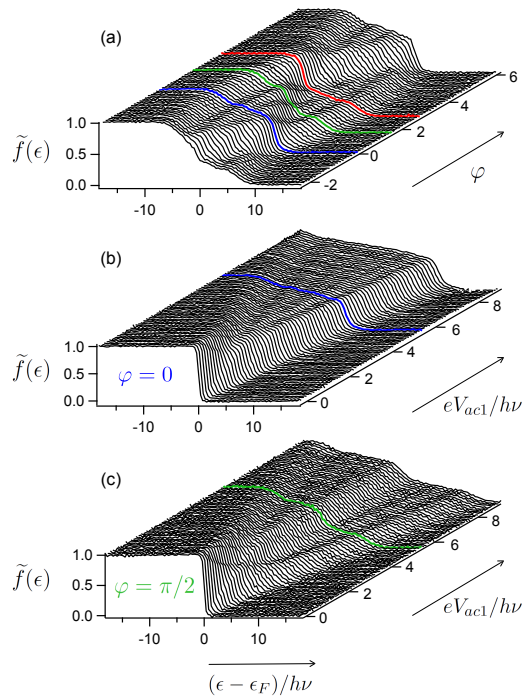}
\end{center}
\vspace{-0.5cm}
\caption{Non-equilibrium distribution functions obtained from numerical deconvolution of the measured differential noise. (a) Noise is measured for $eV_{ac1}=2eV_{ac2}=5.4h\nu$, for phase shifts $\varphi=0$ (blue), $\varphi=\pi/2$ (green) and $\varphi=\pi$ (red). (b) (resp. (c)) Noise is measured for various amplitudes of excitation $V_{ac1}$ and  $V_{ac2}$ keeping $V_{ac1}=2V_{ac2}$, for $\varphi=0$ (resp. $\varphi= \pi/2$).\label{fig2}}
\end{figure}

\emph{Measurements of distribution functions.} Distribution functions in samples driven out of equilibrium by a dc voltage have been obtained by the measurement of the differential conductance in systems where the density of states depends on energy. This occurs with superconducting electrodes, \cite{Pothier,Pierre}, in the presence of dynamical Coulomb blockade \cite{Pierre2,Anthore} or with a quantum dot \cite{Altimiras}. In our case, the differential conductance of the junction is totally voltage-independent and its measurement does not provide any spectroscopic information. However, as we show below, the differential noise does. As with differential conductance measurements, we cannot access the distribution functions of the two contacts separately. We measure the distribution functions that are involved in the transport, which depends only on the voltage difference between the contacts. This can be described by taking one of the contacts at equilibrium while the other one experiences the full time-dependent voltage. For not too small energy $\xi$, $\xi \gg  h\Delta \nu_f, k_BT_{el}$ where $\Delta \nu_f=1.3 \,  \mathrm{GHz}$ is the bandwidth of the noise detection, we show in appendix A that the distribution function $\tilde{f}$ is related to the differential noise $\partial S_2 / \partial eV_{dc}$ by:
\begin{equation}
 \tilde{f}(\epsilon_F+\xi) \simeq  \frac{1}{2}\left(1- \frac{1}{G} \frac{\partial S_2}{\partial eV_{dc}} \right)_{eV_{dc}=\xi}
 \label{eq1}
\end{equation}
\noindent An energy resolution better than $h\Delta\nu,k_B T_{el}$ can be achieved by numerical deconvolution of the noise data, as explained in Appendix A.

\begin{figure}[b]
\begin{center}
\includegraphics[width=0.9\linewidth]{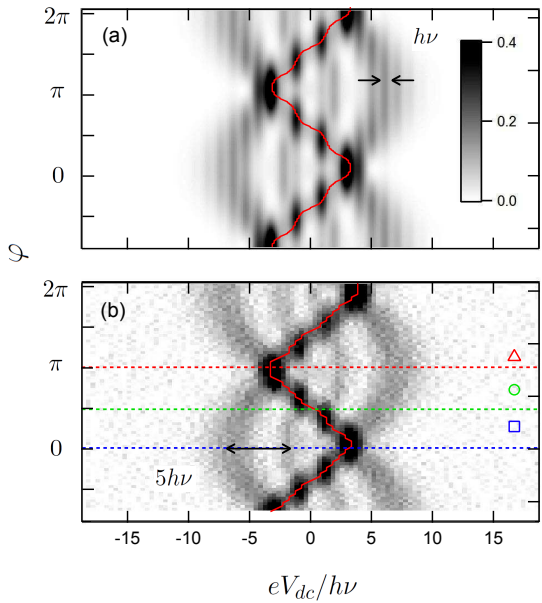}
\end{center}
\vspace{-0.5cm}
\caption{Calculated (a) and measured (b) second derivative of the bi-harmonic photon-assisted noise $\partial^2S_{2,ac}/\partial V_{dc}^2$ as a function of normalized dc bias and phase shift. In both cases $eV_{ac1}=2eV_{ac2}=5.4h\nu$ with $\nu=10$ GHz, and the temperature is $T_{el}= 0.14 \, h\nu/k_B = 75 \, \mathrm{mK}$. Red curves correspond to the calculated (a) and measured (b) minimum of the photon-assisted noise, $\partial S_{2,ac}/\partial V_{dc} =0$. Dashed lines correspond to phase shifts that are used in Fig. \ref{fig4}. Arrows in (a) show the fringes associated with $eV_{dc}=h\nu$. Double arrow in (b) indicates $5h\nu$.}
\label{fig3}\end{figure}

\vspace{0.3cm}\noindent\emph{Distribution function for a time-dependent excitation.} In the presence of a periodic voltage $V_{ac}(t)$ of frequency $\nu$, the electron wave functions acquire an extra phase factor \cite{Tien}:
\begin{equation}
\Xi(t)=\exp \left( \frac{-i}{\hbar} \int_0^t eV_{ac}(t') \, dt' \right)
\label{eq2}
\end{equation}
\noindent Electronic states with energy $\epsilon$ are split into subbands with energies $\epsilon \pm nh\nu$ and spectral weight given by the modulus squared of the Fourier coefficients $c_n$ of $\Xi(t)=\sum_{n=-\infty}^{+\infty}c_n e^{i2\pi \nu nt}$. The corresponding non-equilibrium distribution function is:
\begin{equation}
\tilde{f}(\epsilon)= \sum_{n=-\infty}^{+ \infty} \left|c_n \right|^2 f(\epsilon + n h \nu)
\label{eq3}
\end{equation}
\noindent where $f$ is the equilibrium Fermi-Dirac distribution. For harmonic excitation ($V_{ac2}=0$), $c_n=J_n\left(eV_{ac1}/h\nu\right)$ with $J_n$ the Bessel functions of the first kind. For bi-harmonic excitation:
\begin{equation}
c_n=\sum_{m=-\infty}^{+\infty} J_{n-2m}\left( \frac{eV_{ac1}}{h\nu}\right)J_{m} \left( \frac{eV_{ac2}}{2h\nu}\right) \, e^{-im \varphi}
\label{eq4}
\end{equation}
\noindent The sum in Eq.(\ref{eq4}) expresses the interference involving several absorption/emission processes of photons of frequencies $\nu$ and $2\nu$. This interference depends on the relative phase $\varphi$. Fig. \ref{fig2} shows measured non-equilibrium distribution functions $\tilde{f}$ for different ac excitations. They are obtained from numerical deconvolution of the noise data (see Appendix A). Although bi-harmonic excitation depends on only three parameters $V_{ac1}$, $V_{ac2}$ and $\varphi$, a large class of distribution functions can be realized, which allows to control related physical properties such as the shot noise. For example, taking $V_{ac1}=2V_{ac2}$ creates a distribution function with two steps. The height and width of the steps can be controlled by tuning the phase shift (see Fig. \ref{fig2}(a)) or the amplitude of the ac excitation (see Fig. \ref{fig2}(b-c)). We show in the following that this distribution minimizes the shot noise for a given amplitude $V_{ac1}$. Making the spectroscopy of a system with discrete levels has been performed in solid state qubits with harmonic \cite{Berns2} and biharmonic \cite{Bylander} excitation. In such systems, one can directly measure the population of the levels in the presence of the excitation. In our case, the noise, i.e. the variance of the fluctuations of the populations, provides the spectroscopic information.

\vspace{0.3cm}\noindent\emph{Noise minimization.} In the following we show how controlling the distribution function via the shape of the exciting waveform allows to reduce the shot noise in the tunnel junction. The current noise of a coherent conductor biased by a time-dependent, periodic voltage has been calculated for a pure sine wave excitation \cite{Lesovik,Pedersen}. For a tunnel junction and an arbitrary periodic excitation, we obtain:
\begin{equation}
S_{2,ac}(eV_{dc})=\sum_{n=-\infty}^{+\infty} \left|c_n \right|^2 \, S_2^{(0)}(eV_{dc}+nh\nu)
\label{eq5}
\end{equation}
\noindent where $S_2^{(0)}(eV)=GeV \coth \left(eV/2k_BT_{el} \right)$ is the Johnson-Nyquist equilibrium noise, and $c_n$ are given by Eq. (\ref{eq4}). In the case of a harmonic excitation, one observes features on $S_{2,ac}(eV_{dc})$ at bias $eV_{dc}=nh \nu$ with $n$ integer (discontinuities of $dS_2/dV$ rounded by the finite temperature and detection bandwidth) \cite{Schoelkopf,GR1}. For bi-harmonic excitation the interferences between multi-photon assisted processes at frequency $\nu$ and $2 \nu$ induce interference fringes on a larger scale. We show this additional complexity in interference pattern for $V_{ac1}=2V_{ac2}$ in Fig.\ref{fig3}(b), where the  second derivative of the noise, $\partial^2 S_{2,ac}/\partial eV_{dc}^2$, is plotted. The choice $V_{ac1}=2V_{ac2}$ has been motivated by a numerical calculation described in Appendix B. Interference pattern in the  $(eV_{dc},\varphi)$ space exhibits fringes with a fringe spacing $\simeq 5h\nu$ (see Fig.\ref{fig3}(b)), in agreement with numerical calculations using Eqs. (\ref{eq4}) and (\ref{eq5}), see Fig. \ref{fig3}(a), whereas the substructure at $h\nu$ is almost washed out by thermal broadening. Red curves in Fig.\ref{fig3} correspond to the calculated (a) and measured (b) $eV_{dc}$ value at which the photon-assisted noise is minimal ($\partial S_{2,ac}/\partial eV_{dc}=0$). It exhibits steps at $eV_{dc}=\pm h\nu$ and $eV_{dc}=\pm 3h\nu$.

The appearance of fringes at a scale larger than $h\nu$ is similar to what is observed when systems with discrete spectrum (such as qubits) are driven by a large amplitude signal \cite{Berns1,Berns2}: to the fringes due to individual photon resonances, characterized by the energy scale $h\nu$, are superimposed fringes with larger characteristic scale corresponding to St\"uckelberg oscillations. The latter may persist even if the $h\nu$ pattern is lost (as when the excitation frequency is smaller than the inverse decoherence time) and are a direct consequence of quantum coherence. They are caused by interferences involving several possible Landau-Zener transitions between adjacent levels, with a phase set by the time-dependence of the energy levels. In a similar way, in our case, two contacts with time-dependent chemical potentials are coupled by tunneling. The phase acquired by the electron-hole pairs involved in the transport mechanism depends on the time-dependence of the voltage. The probability to cross the barrier involves interferences between several processes, which results in the St\"uckelberg-like oscillations we observe \cite{Levitov_com}. This behavior is generic for driven quantum systems and is a part of the more general effect of Ramsey multiple-time-slit interferences \cite{Akkermans}.

\begin{figure}[t]
\begin{center}
\includegraphics[width=0.75\linewidth]{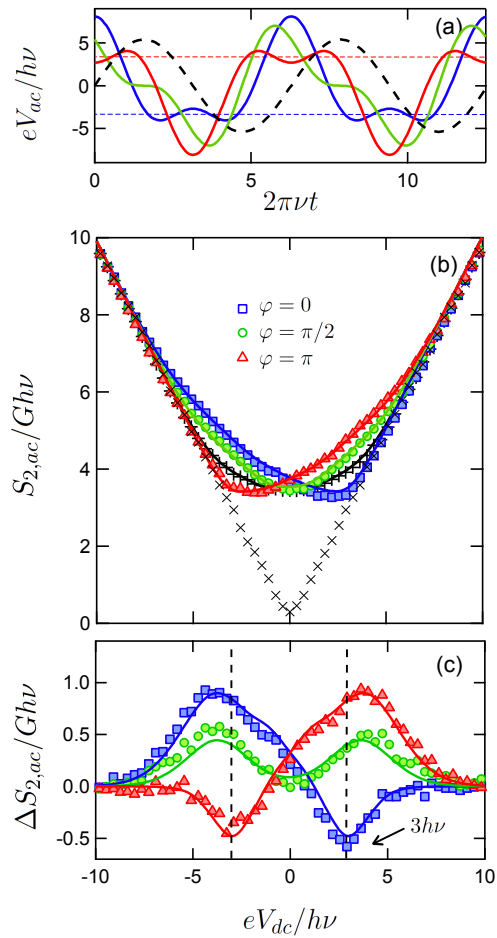}
\end{center}
\vspace{-0.5cm}
\caption{(a) Shape of the ac excitation with the different phase shifts: $\varphi=0$ (blue), $\varphi=\pi/2$ (green) and $\varphi=\pi$ (red). Dashed line: sine wave of the same amplitude. (b) Normalized bi-harmonic photon-assisted noise $S_{2,ac}/Gh\nu$ vs normalized dc bias for $eV_{ac1}=5.4h\nu$. Blue square, green circle, red triangle symbols: data for $eV_{ac2}=2.7h\nu$ and phase shifts $\varphi= 0, \pi/2, \pi$. Plus symbols ($+$): data for $V_{ac2}=0$, i.e., pure sine wave excitation. Cross symbols ($\times$): data for $V_{ac1}=V_{ac2}=0$, i.e., shot noise without any ac excitation. Solid lines: theoretical predictions, Eq. (5,6). (c) Difference between bi-harmonic and  mono-harmonic photon-assisted noise $\Delta S_{2,ac}(V_{dc})=S_{2,ac}(V_{dc},V_{ac1},V_{ac2})-S_{2,ac}(V_{dc},V_{ac1},V_{ac2}=0)$.\label{fig4}}
\end{figure}

We have calculated numerically the set of parameters $(eV_{dc},eV_{ac2},\varphi)$ minimizing the photon-assisted noise $S_{2,ac}$ for a given $eV_{ac1}$ and temperature $T_{el}$, see Appendix B. The optimal dc voltage is zero only for $\varphi=\pi/2$, which corresponds to the existence of a symmetry in the waveform: for each positive value of $V_{ac}(t)$ there is a symmetric, negative value (green curve on Fig.\ref{fig4}(a)). Thus the ac voltage explores positive and negative values in the same way. When this symmetry is lost there is no reason for the noise to reach its minimum at $V_{dc}=0$. For experimental parameters $T=0.14 \, h\nu$ and $eV_{ac1}=5.4 \, h\nu$, we obtain that optimal values are $eV_{ac2}=eV_{dc}= 2.4 \, h\nu$ and $\varphi=0$. For $\varphi=\pi$, the waveform is reversed (see Fig.\ref{fig4}(a)) and the minimum occurs at the opposite value of $V_{dc}$.  Fig.\ref{fig4}(b) shows noise measured for $eV_{ac2}= 2.7 \, h\nu$ (i.e., close to optimal) for $\varphi=0$ (blue), $\pi/2$ (green) and $\pi$ (red). All the data (symbols) are very well fitted by the theory (solid lines). One observes that for a given waveform, the noise is minimal at $eV_{dc}=0$ only for $\varphi=\pi/2$, as expected, whereas the minima for $\varphi=0$ and $\pi$ occur at opposite values of $eV_{dc}=\pm 2.3 \, h\nu$ in agreement with the numerical result.

The black curve on Fig. \ref{fig4}(a) (plus symbol $+$) shows the noise for $V_{ac2}=0$. There is a clear region of $V_{dc}$ where it is \emph{above} the red or blue curve, which correspond to $V_{ac2}\neq 0$. We have emphasized this result by plotting on Fig. \ref{fig4}(c) the difference $\Delta S_{2,ac}(V_{dc})=S_{2,ac}(V_{dc},V_{ac1},V_{ac2})-S_{2,ac}(V_{dc},V_{ac1},V_{ac2}=0)$ between the noise under bi-harmonic and mono-harmonic excitations, which can be negative. This proves that the addition of the excitation at frequency $2\nu$ may \emph{reduce} the noise. It is also noticeable that the noise under bi-harmonic excitations drops below the absolute minimum of the noise with mono-harmonic excitation, which occurs at zero bias, in agreement with our numerical simulations (see Appendix B). This is a purely quantum effect.

Noise has been predicted to be minimal when the excitation is a sequence of Lorentzian peaks of a quantized area $\int_0^T eV(t) \, dt=Nh$ \cite{Levitov1,Levitov2}. Such an excitation does not add more noise than its dc voltage alone. In other words, the noise as a function of the dc voltage has minima for quantized values of $V_{dc}$. This property seems to be valid for many ac waveforms at zero temperature \cite{Vanevic}, including the bi-harmonic excitation (see Appendix C). We observe that this is no longer the case at finite temperature for the biharmonic excitation (data not shown), in agreement with numerical calculations (see Appendix C). Let us now consider the difference in the noise for two excitations at the same frequency. Obviously, it should have extrema for the same quantized values of $V_{dc}$ at zero temperature. As shown in Fig. \ref{fig4}(b), this property seems to survive at finite temperature if we consider the difference between the mono-harmonic and bi-harmonic excitations $\Delta S_{2,ac}$, which has minima at $\pm 3 h\nu$.

It is interesting to remark that the waveform we found that minimizes the noise for a given $V_{ac1}$ at finite temperature is not close to Lorentzian, but corresponds almost to the first two harmonics of a Lorentzian with a dc offset (see Fig. \ref{fig1}B). The Lorentzian pulses are optimal only if we consider the noise at zero frequency, zero temperature and integer values of $eV_{dc}/h\nu$. They are no longer optimal if we work at finite detection frequency, finite temperature or non-integer value of $eV_{dc}/h\nu$, see Appendix C.

\vspace{0.3cm}\noindent\emph{Conclusion.} We have observed the effect of bi-harmonic illumination on the non-equilibrium current noise in a tunnel junction. We have measured the low frequency shot noise of the junction while varying the shape of the ac excitation and showed that from these measurements we can determine the out-of-equilibrium distribution function induced by the excitation. This opens the way of engineering the waveform of an ac signal to control the out-of-equilibrium distribution function of the electrons in a mesoscopic conductor, thus modifying its physical properties. We have demonstrated this ability by reducing the shot noise in a tunnel junction irradiated at frequency $\nu$ by adding another coherent irradiation at frequency $2\nu$ of controlled amplitude and phase. Such a procedure may be used in many situations. For example, it may be used to dynamically control the amplitude of the critical current of a superconductor / normal metal / superconductor tunnel junction \cite{Francesca}, or even reverse it as with a dc current \cite{Baselmans,DP}. This would  realize a Josephson junction that can be switched from $0-$ to $\pi-$state dynamically, an interesting device in the context of quantum computation.

\vspace{0.3cm}\noindent\emph{Acknowledgements.} We are very grateful to Lafe Spietz for providing us with the sample and to Leonid Levitov for many stimulating discussions. We thank Marco Aprili and Sophie Gu\'eron for fruitful discussions. This work was supported by ANR-11-JS04-006-01 and CERC "Quantum Signal Processing".

\section{Appendix A: Tunneling spectroscopy of distribution functions}

\noindent The quantity we measure is:
\begin{equation}\label{eqA0}
\Delta I^2(eV_{dc})= \int_{-\infty}^{+\infty} S_{2}(eV_{dc},h\nu') \, |H(\nu')|^2 \, d\nu'
\end{equation}
\noindent where $S_{2}(eV_{dc},h \nu')$ is the spectral density of current fluctuations at frequency $\nu'$ and $H(\nu')$ the frequency response of the bandpass filter, of width $\Delta \nu_f$ and central frequency $\nu_f$ (in our experiment, $\nu_f=1.15$ GHz, $\Delta\nu_f=1.3$ GHz). $S_{2}(eV_{dc},h \nu')$ depends on the ac excitation $V_{ac}(t)$. For a tunnel junction with energy-independent transmissions, $S_2(eV_{dc},h \nu')$ is simply:
\begin{equation}
\begin{split}
S_{2}(eV_{dc},h \nu')=& \, \frac{1}{2} \left(S_{2}^{(0)}(eV_{dc}+h \nu') \right. \\
& \hspace{1cm}  \left.+S_{2}^{(0)}(eV_{dc}-h \nu')\right)
\end{split}
\label{eqA1}
\end{equation}
With $S_{2}^{(0)}(eV_{dc})=S_{2}(eV_{dc},0)$ the zero-frequency noise spectral density, given by:
\begin{equation}
\begin{split}
S_{2}^{(0)}(eV_{dc})= & \, G  \int_{-\infty}^{+\infty} \left[ f_L(\epsilon) (1-f_R(\epsilon)) \right.\\
& \hspace{1cm} \left.  + f_R(\epsilon)(1-f_L(\epsilon))\right] \, d\epsilon
\end{split}
\label{eqA3}
\end{equation}
Here $f_L$ (respectively $f_R$) is the energy distribution function of electrons in the left (resp. right) contact. In the presence of both dc and ac bias, we can without loss of generality consider that the dc bias is applied to the left reservoir while the ac voltage is applied to the right one. Thus $f_L$ is the Fermi-Dirac distribution $f$ with a shifted electrochemical potential, $f_L(\epsilon)=f(\epsilon - eV_{dc})$, whereas $f_R=\tilde{f}$ is the non-equilibrium distribution function we wish to measure. Defining the difference in the noise with and without ac excitation, $M(eV_{dc})=\Delta I^2(eV_{dc},eV_{ac}\neq0)-\Delta I^2(eV_{dc},eV_{ac}=0)$, we obtain the following convolution:
\begin{equation}\label{eqA4}
\frac{\partial M}{\partial eV_{dc}}(eV_{dc})=  \int_{-\infty}^{+\infty} K\left(eV_{dc}-\epsilon \right)\,\left( \tilde{f}(\epsilon)-f(\epsilon)\right) \, d\epsilon
\end{equation}
\noindent with a kernel:
\begin{equation}\label{eqA4b}
K(\epsilon)= -\frac{G}{h} \, \int_{-\infty}^{+\infty} |H(\nu')|^2 \,  \frac{\partial f}{\partial \nu'}(h \nu'-\epsilon) \, d\nu'
\end{equation}
Thus, the non-equilibrium function $\tilde{f}$ can be calculated using the Fourier transform $FT$:
\begin{equation}\label{eqA5}
\tilde{f}(\epsilon)=f(\epsilon)+FT^{-1}\left\{\frac{FT[\frac{\partial M}{\partial eV_{dc}}]}{FT[K]}\right\}(\epsilon)
\end{equation}
\noindent In the limit $\xi \gg  h\nu_F,h \Delta \nu_f, k_BT_{el}$, using $K(\epsilon)=-4G \Delta \nu_f \, \delta(\epsilon-\epsilon_F)$, (\ref{eqA5}) reduces to:
\begin{equation}
 \tilde{f}(\epsilon_F+\xi) \simeq  \frac{1}{2}\left(1- \frac{1}{G} \frac{\partial S_2}{\partial eV_{dc}} \right)_{eV_{dc}=\xi}
 \label{eqA6}
\end{equation}
In our experiment, $h\nu_f/k_B=50$ mK, $h\Delta\nu_f/k_B=56$ mK and $T_{el}=70$ mK.

\begin{figure}[htb!]
\begin{center}
\includegraphics[width=0.8\linewidth]{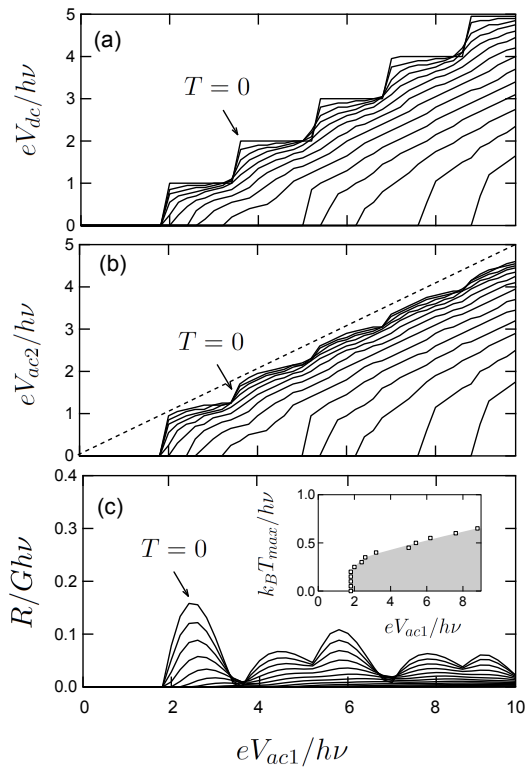}
\end{center}
\vspace{-0.5cm}
\caption{(a) Optimal value of the reduced dc voltage $eV_{dc}/h\nu$. Optimal value of the reduced amplitude $eV_{ac2}/h\nu$ at frequency $2\nu$. (c): Noise reduction $R= S_{2} (V_{dc}=0,V_{ac1},V_{ac2}=0) -S_{2}(V_{dc}^\star,V_{ac1},V_{ac2}^\star)$ between the mono-harmonic photon-assisted noise at zero dc bias and the optimal bi-harmonic one with the same ac amplitude $V_{ac1}$ at frequency $\nu$. (a), (b) and (c) are plotted as a function of the reduced amplitude $eV_{ac1}/h\nu$ at frequency $\nu$, for various reduced temperature $k_BT_{el}/h\nu$ ranging from 0 (blue line) to 0.65 by step of 0.05. Inset: temperature above which the noise cannot be reduced by bi-harmonic excitation.\label{fig5}}
\end{figure}

\section{Appendix B: Optimization of the bi-harmonic photon-assisted noise at finite temperature}

\begin{figure}[htb]
\begin{center}
\includegraphics[width=0.7\linewidth]{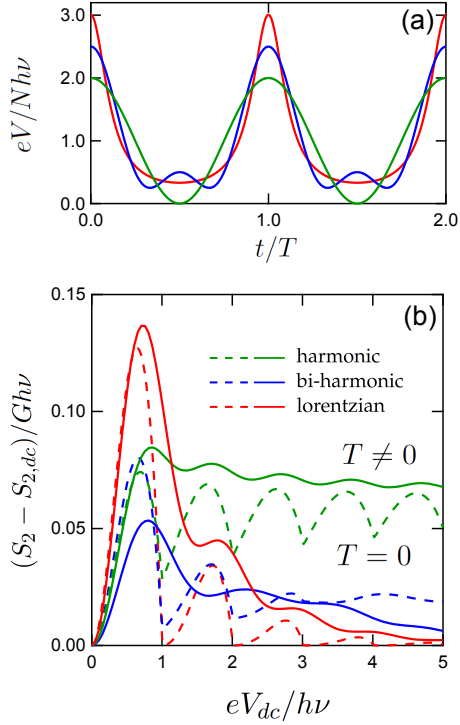}
\end{center}
\vspace{-0.5cm}
\caption{(a) $T$-Periodic sequence of Lorentzian pulses of width $\tau=\mathrm{ln}\,2/2 \pi \, T$ (red line, $V_L(t)$) and its harmonic (green line, $V_1(t)$) and bi-harmonic (blue line, $V_2(t)$) approximations. (b) Noise difference $S_{2}-S_{2,dc}$ for the different waveforms. Dashed lines correspond to zero temperature whereas solid lines correspond to our experimental temperature, $k_BT=0.14\,h\nu$.\label{fig6}}
\end{figure}

\noindent The photon-assisted noise in the tunnel junction depends on the shape of the ac excitation. In the case of bi-harmonic excitation, we determine numerically the set of parameters $(eV_{dc}^{\star},eV_{ac1}^{\star},eV_{ac2}^{\star},\varphi^{\star})$ which minimize the noise at temperature $T_{el}$. At each temperature, the noise spectral density is calculated for $100\times100\times100\times100$ different values of $\left(eV_{dc}/h\nu,eV_{ac1}/h\nu,eV_{ac2}/h\nu,\varphi\right)$ in the range $[-5,5] \times[0,10] \times[0,5] \times [0,\pi]$. Let us suppose we excite at frequency $\nu$ with an amplitude $V_{ac1}$ and we want to minimize the low frequency shot noise. Figures \ref{fig5}(a) and \ref{fig5}(b) show respectively how to choose $V_{dc}$ and $V_{ac2}$ to reach this goal, as a function of $V_{ac1}$ and for various $T_{el}$. The obtained noise reduction, $R= S_{2} (V_{dc}=0,V_{ac1},V_{ac2}=0) -S_{2}(V_{dc}^\star,V_{ac1},V_{ac2}^\star)$, is plotted in Fig. \ref{fig5}(c). It appears that one always have $V_{dc}^\star \simeq V_{ac2}^\star$. For example, for our experimental parameters $k_BT_{el}=0.14h\nu$ and $eV_{ac1}=5.4h\nu$, the optimum is $eV_{dc}=2.38 h\nu$, $eV_{ac2}=2.4 h\nu$ and $\varphi=0$ (or the opposite $V_{dc}$ for $\varphi=\pi$). Adding dc voltage and ac voltage at frequency $2\nu$ allows to reduce the noise below that with no dc bias and the same excitation at frequency $\nu$, by an amount $R=0.04 \, Gh \nu$ (or in terms of noise temperature, by 20 mK). We observe this effect, see Fig.4(b) of the letter: the minimum of $S_2(V_{dc},V_{ac1},V_{ac2}^\star)$ drops below $S_2(V_{dc}=0,V_{ac1},V_{ac2}=0)$ for $V_{dc}\sim V_{dc}^\star$.

For a given $V_{ac1}$, there is a temperature $T_{max}(eV_{ac_1})$ above which the optimal point does not exist anymore: $V_{dc}^{\star}=V_{ac2}^{\star}=0$, see inset of Fig. \ref{fig5}(c). Above that temperature, adding a second harmonic will never reduce the noise. In particular, for  $eV_{ac1}<2h\nu$ it is never possible to reduce the noise with a bi-harmonic-excitation. In our experiment, $T_{max}\simeq250\,\mathrm{mK}>T_{el}$, so the addition of the sine wave at frequency $2\nu$ may lead to a reduction of the noise, as we observe.

\vspace{1cm}
\section{Appendix C: photon-assisted noise at finite temperature for various waveforms}

\noindent We consider the three waveforms shown on Fig.\ref{fig6}(a): $V_L(t)$ is the Lorentzian shape of width $\tau=\mathrm{ln}\,2/2 \pi \, T$ (this value is chosen to have the same first two harmonics as the one we have chosen in our experiment); $V_1(t)=V_{dc}[1+\cos(2\pi \nu t)]$ is the same Lorentzian waveform truncated to the same dc and first harmonic; $V_2(t)=V_{dc}[1+\cos(2\pi \nu t)+0.5\cos(4\pi \nu t)]$ is again the same Lorentzian waveform but truncated to dc and first two harmonics. We show on Fig.\ref{fig6}(b) the numerical difference in the noise, $S_2-S_{2,dc}$ between ac+dc excitation and dc only excitation for these three waveforms. At zero temperature (dashed lines), there is a sharp minimum for each integer value of $eV_{dc}/(h\nu)$ for the three waveforms. The Lorentzian reaches zero and the bi-harmonic is better than the mono-harmonic. For $eV_{dc}<h\nu$ the Lorentzian is the worst. At finite temperature (solid lines, $k_BT=0.14\,h\nu$ as in the experiment), none of the waveforms minimize the noise at quantized values of $eV_{dc}/h\nu$.


\vspace{-5mm}
\begin{thebibliography}{99}
\vspace{-5mm}
\bibitem{Levitov1} L.S. Levitov, H.W. Lee and G.B. Lesovik, J. Math. Phys. \textbf{37}, 4845 (1996).
\bibitem{Levitov2} D.A. Ivanov, H.W. Lee and L.S. Levitov, Phys. Rev. B \textbf{56}, 6839-6850 (1997).
\bibitem{Klich} J. Keeling, I. Klich and L. S. Levitov, Phys. Rev. Lett. \textbf{97}, 116403 (2006).
\bibitem{Degiovanni} C. Grenier, R. Herv\'e, E. Bocquillon, F. D. Parmentier, B. Pla\c{c}ais, J.-M. Berroir, G. F\`eve, P. Degiovanni, New J. of Phys. \textbf{13}, 093007 (2011).
\bibitem{Ambrumenil} N. d' Ambrumenil and B. Muzykantskii, Phys. Rev. B\textbf{71}, 045326 (2005).
\bibitem{Vanevic} M. Vanevi\'c, Y.V. Nazarov and W. Belzig, Phys. Rev. B\textbf{78}, 245308 (2008).

\bibitem{Lesovik} G.B. Lesovik and L.S. Levitov, Phys. Rev. Lett. \textbf{72}, 538 (1994).
\bibitem{Schoelkopf} R.J. Schoelkopf, A.A. Kozhevnikov and D.E. Prober and M.J. Rooks, Phys. Rev. Lett. \textbf{80}, 2437 (1998).
\bibitem{Reydellet} L.-H. Reydellet, P. Roche, D. C. Glattli, B. Etienne, and Y. Jin, Phys. Rev. Lett. \textbf{90}, 176803 (2003).
\bibitem{GR1} J. Gabelli and B. Reulet, Phys. Rev. Lett. {\bf 100},  026601 (2008).

\bibitem{Lafe} L. Spietz, K.W. Lehnert, I. Siddiqi and R.J. Schoelkopf, Science \textbf{300}, 1929 (2003).

\bibitem{Pothier} H. Pothier, S. Gu\'eron, Norman O. Birge, D. Esteve and M.H. Devoret, Phys. Rev. Lett. \textbf{79}, 3490 (1997).
\bibitem{Pierre} F. Pierre, A. Anthore, H. Pothier, C. Urbina, and D. Esteve, Phys. Rev. Lett. \textbf{86}, 1078 (2001).
\bibitem{Pierre2} F. Pierre, H. Pothier, P. Joyez, Norman O. Birge, D. Esteve, and M. H. Devoret, Phys. Rev. Lett. \textbf{86}, 1590 (2001).
\bibitem{Anthore} A. Anthore, F. Pierre, H. Pothier, and D. Esteve, Phys. Rev. Lett. \textbf{90}, 076806 (2003).
\bibitem{Altimiras} C. Altimiras, H. le Sueur, U. Gennser, A. Cavanna, D. Mailly and F. Pierre, Nature Physics 6, 34 (2010).
\bibitem{BuBlan} Y.M. Blanter and M. B\"uttiker, Phys. Rep. \textbf{336}, 1 (2000).

\bibitem{Tien} P.K. Tien and J.P. Gordon, Phys. Rev. \textbf{129}, 647 (1963).

\bibitem{Berns2} D.M. Berns, M.S. Rudner, S.O. Valenzuela, K.K. Berggren, W.D. Oliver, L.S. Levitov and T.P. Orlando, Nature \textbf{455}, 51 (2008).
\bibitem{Bylander} J. Bylander, M.S. Rudner, A.V. Shytov, S.O. Valenzuela, D.M. Berns, K.K. Berggren, L.S. Levitov and W.D. Oliver, Phys. Rev. B\textbf{80}, 220506(R) (2009).

\bibitem{Pedersen} M.H. Pedersen and M. B\"uttiker, Phys. Rev. B \textbf{58}, 12993 (1998).
\bibitem{Berns1} D.M. Berns, W.D. Oliver, S.O. Valenzuela, A.V. Shytov, K.K. Berggren, L.S. Levitov and T.P. Orlando, Phys. Rev. Lett. \textbf{97}, 150502 (2006).
\bibitem{Levitov_com} L.S. Levitov, private comm.
\bibitem{Akkermans} E. Akkermans and G.V. Dunne, Phys. Rev. Lett. \textbf{108}, 030401 (2012).
\bibitem{Francesca} F. Chiodi, M. Aprili and B. Reulet, Phys. Rev. Lett. \textbf{103}, 177002 (2009).
\bibitem{Baselmans} J. J. A. Baselmans, A. F. Morpurgo, B. J. Van Wees, T. M. Klapwijk, Nature \textbf{397}, 43 (1999).
\bibitem{DP} D. Prober, private comm.
\end{thebibliography}
\end{document}